\documentclass[12pt]{article}
\usepackage{graphicx,color} 
\usepackage{amsmath}  
\usepackage{amssymb}
\usepackage{epstopdf}

\def\be{\begin{eqnarray}}
\def\ee{\end{eqnarray}}

\def\ket{{\rangle}}

\def\ds{\displaystyle}

\def\pb{{\bf p}}
\def\qb{{\bf q}}
\def\rb{{\bf r}}
\def\ie{{\it i.e.\;}}
\def\ni{\noindent}

\begin{document}

\title{Higgs particles interacting via a scalar Dark Matter field}

\author{Yajnavalkya Bhattacharya \\ 
\and
 Jurij W. Darewych \\ York University, Toronto, Canada \\ darewych@yorku.ca}    

\maketitle

\begin{abstract}
  We study a system of two Higgs bound state, interacting  via a real scalar Dark Matter mediating field,  without imposing $Z_2$ symmetry on the DM sector of the postulated Lagrangian. The variational method in the Hamiltonian formalism of QFT is used to derive relativistic wave equations for the two-Higgs system, using a truncated Fock-space trial state. Approximate solutions of the 2-body relativistic coupled integral equations are presented, and conditions for the existence of Higgs bound states is examined in a  broad parameter space of  DM mass and coupling constants. \\
  
\end{abstract}

\section{Introduction}

Astrophysical observations of gravitational effects at all length scales of the universe, including our Milky Way galaxy, indicate the existence of Dark Matter (DM) \cite{1,2,201}.  To date, DM has been observed only through its gravitational influence such as rotation curves of galaxies, mismatch of estimated mass and luminous matter in galaxy clusters, lensing of galactic Supernovae of Type Ia and Cosmic Microwave Background images, etc. A major open question is the nature of DM, since to date, there are no direct measurements of DM's electroweak, or any non-gravitational, interactions.  
\par 
One conjecture is that DM is made up non-baryonic particles called WIMPs (Weakly Interacting Massive Particles). Experimental efforts at detecting DM, including WIMPs,  are ongoing \cite{202}. The literature on this subject is extensive. A brief, useful overview of DM and efforts to detect them is given in ref. \cite{3}.
\par
An outstanding question is how DM interacts with Standard Model particles, including the Higgs particles. There have been a number of theoretical models that address this problem (see, for example, references \cite{5,6,7} and citations therein). Recently, Petersson et al. have proposed a Supersymmetric generalization of the Standard Model in which the Higgs boson can disintegrate into a photon and DM particles \cite{8}.      
\par
The possibility of Higgs bound states (``Higgsonium") has been examined well before the experimental detection of the Higgs 
boson (cf. references \cite{9} to \cite{13} ).  In these prior studies, the domain of the Higgs mass and coupling strength for which Higgsonium binding might occur, were investigated. These 2-Higgs bound-state results (due to Higgs self coupling) are somewhat moot, now that the Standard Model Higgs mass has been observed to be $125.7$ GeV.

\vskip .2cm
The present work is concerned with the interactions of Higgs particles with DM particles. 
In particular we shall examine the possibility of two-body Higgs bound states due a mediating  DM  field. To our knowledge, this is the first investigation of Higgs relativistic bound state formation, using a general DM Lagrangian, via a  singlet scalar DM channel. 
\section{The model}
In this work the Dark Matter is assumed to be a spinless,  singlet, massive scalar field $\phi$,   of mass $\mu$, coupled with the
Higgs field $\chi$, which is taken to be the Higgs sector of the Standard Model. Models with scalar singlet DM have been studied previously (see for e.g. references \cite{15} \cite{16}). 						
In this study,  we shall consider the interaction of the Higgs particles with DM only. 
The interaction terms will be taken to be of a general form consistent with  classical stability and renormalizability, 
that is the  postulated Lagrangian density of this model is ($\hbar=c=1$)

\be  \label{1.2}
\begin{split}
	\mathcal L \,= \,\frac{1}{2} \partial^{\nu} \phi \, \partial_{\nu} \phi - \frac{1}{2}\mu^{2} \phi^{2}
	- \sigma \phi^3 - \kappa \, \phi^4   \\ 
	+ \, \frac{1}{2} \partial^{\nu} \chi \, \partial_{\nu} \chi - \frac{1}{2} m^{2} \chi^{2} 
	- \lambda \, v \, \chi^{3} - \frac{1}{4}\lambda \, \chi{^4} \\
 - \, g_1 \, \chi \, \phi^2 - \eta_1 \chi^2 \phi^2 - \eta_2 \chi \phi^3 - \eta_3 \chi^3 \phi - g_2 \, \chi^2 \, \phi ,
 \end{split}
\ee
where $\kappa, \lambda$, $g_1$, $g_2$, $v$ and $\eta_j \;(j= 1,2,3)$ are positive coupling constants;\, $\lambda, \kappa, \eta_j$ being dimensionless, and $v, \sigma$, $g_j$, having dimensions of mass. Aside from the Higgs sector coupling constants $\lambda$ and $v$, 
all the others are presently unknown and will be treated to be adjustable. 

Note that we have not included  a $g \, \chi \phi$ term in our postulated Lagrangian (\ref{1.2}) becase such a mixing term would contain
a coupling constant $g$ with dimensions of $({\rm mass})^2$, which is unusual and would allow for instability of the DM.

 The proposed Lagrangian (\ref{1.2}) does allows for the decay of a very heavy single DM particle into two Higgs, provided that 
$m_{DM}\,\ge\,2\, m_{Higgs}$, where $m_{DM} \equiv \mu$, and $m_{Higgs} \equiv m$ are the DM and Higgs masses respectively.

 In order to decrease the coupling-constant parameter space we shall set $\sigma = \kappa = \eta_i = 0$ for this paper.

 It is shown below that in the domain where the effective Higgs-DM dimensionless coupling constant varies from $0$--$2$,
two Higgs bound states are formed only if $m_{DM}\, \leq \, m_{Higgs}$, which excludes DM decay into the Higgs.

\par
Models in which the DM is represented by a scalar with an assumed $Z_2$ symmetry, have been investigated by many authors, e.g. \cite{15, 17, 18, 21}. Our Lagrangian, however, does not impose a $Z_2$ symmetry,  since the DM particle is represented by a spinless real scalar -- thus allowing for trilinear terms, such as $g_{2} \chi^{2} \phi$, in which the dimensionful coupling constant is similar to that of the Higgs sector of the DM. 
\par
Our choice is motivated by the fact that the popularity of the 
$Z_{2}$ symmetry has largely resulted in the avoidance of studying the simplest case of a real scalar DM with no additional internal degrees of freedom. In an article on scalar DM, Rodejohann and Yaguna \cite{22} have noted that ``..use of discrete symmetries is questionable not only due to its lack of motivation, but also because they are expected to be broken by gravitational effects at the Planck scale, including dark matter decay and likely destroying the feasibility of such models. That is why it is often implicitly assumed that such a discrete symmetries are actually the remnants of additional gauge or flavor symmetries present at a higher scale, thereby delegating the problem to a framework larger than the model under consideration." In another recent paper, Cirelli \cite{23} notes that ``The ``stabilization symmetry" has become such a household tool for the model builder that
often he/she does not even spend time arguing about it: when in a hurry, just say that you add a $Z_2$ symmetry and move on." 
\section{Quantization and Hamiltonian Formalism}
Upon canonical quantization (in the interaction picture) the classical fields $\phi, \chi$ become operators
\begin{equation} \label{1.9}
\phi (x) = \int \frac{d^3p}{\sqrt{(2\pi)^3 \, 2 \omega(p,\mu)}}\big[d(\pb) \, e^{-i p_\mu \cdot x} + d^\dag(\pb) \,  
e^{i p_\mu \cdot x}\big], 
\end{equation}
\be \label{1.10}
\chi (x) = \int \frac{d^3p}{\sqrt{(2\pi)^3 \, 2 \omega(p,m)}}\big[h(\pb) \, e^{-i p_m \cdot x} + h^\dag(\pb) \, 
 e^{i p_m \cdot x}\big], 
\ee
where ~~ $\omega(p,m)=\sqrt{\pb^2 + m^2}$, ~~ $\omega(p,\mu) = \sqrt{\pb^2 + \mu^2}$, 
~~ $x = (t,\rb),~~p_m = \left[ \omega(p,m), \pb \right]$ ~ and $p_\mu= \left[\omega(p,\mu), \pb \right]$. 
The DM and Higgs operators $d^\dag, d$ and $h^\dag, h$ satisfy the usual commutation rules,
\be \label{1.11}
[d(\pb),d^\dag(\qb)] = \delta^3(\pb - \qb) ~~~{\rm and} ~~~[h(\pb),h^\dag(\qb)] = \delta^3(\pb - \qb),
\ee
and all others vanish. 

\par
In the Hamiltonian formalism of QFT, the equations to be solved are 
\be \label{1.11a}
{\hat{P}}^{\beta }|\Psi {\rangle }=Q^{\beta }|\Psi {\rangle },
\ee 
where ${\hat{P}}^{\beta }=({\hat{H}},{\hat{\mathbf{P}}})$ and $Q^{\beta }=(E,\mathbf{Q})$ are the
energy-momentum operator and corresponding eigenvalues. The $\beta = 0$ (energy) component of equation 
(\ref{1.11a}) is generally impossible to solve and this applies to the present model.
Thus, approximate solutions need to be obtained.      
We shall use the variational method, which is applicable to strongly coupled systems. This method is based on the principle
\be \label {1.11b}
\langle \delta \Psi _{trial}|{\hat{H}}-E|\Psi _{trial}\rangle _{t=0}=0 , 
\ee
where $\hat{H}$ is normal ordered, and $|\Psi _{trial}\rangle $ is a suitable trial state. The subscript $t$=0 in equation (\ref{1.11b}) indicates transformation to the Schr\"{o}dinger picture, which is convenient for bound states. 

\section{Trial state and channel equations}
For a system of two Higgs particles interacting via the DM field, the simplest trial state that yields tractable, non-trivial results is
\be \label{12A}
|\psi_{trial}\ket=\int d{\bf p_{1}}  d{\bf p_{2}} F_{1}({\bf p_{1}},{\bf p_{2}})\, h^{\dagger}({\bf p_{1}})h^{\dagger}({\bf p_{2}})\,|0\ket +\nonumber\\
 \int  d{\bf p_{1}}\, d{\bf p_{2}}\, d{\bf p_{3}} \,F_{2}({\bf p_{1}},{\bf p_{2}},{\bf p_{3}})
	h^{\dagger}({\bf p_{1}})h^{\dagger}({\bf p_{2}})d^{\dagger}({\bf p_{3}}) |0 \ket ,
\ee
where $h$ denotes Higgs and $d$ Dark Matter, and $F_{i}$, $(i=1,2)$ are variational channel wave functions to be determined.
We shall work in the rest frame in which ${\hat{\mathbf{P}}} |\psi_{trial}\ket = 0$, which implies that ${\bf p_1 + \bf p_2} = 0$ 
in $F_1(\pb_1,\pb_2)$ and $\pb_1 + \pb_2 + \pb_3 = 0$ in $F_1(\pb_1,\pb_2,\pb_3))$. 
Substituting the expression ({\ref{12A}) into equation (\ref{1.11b}), evaluating the indicated matrix elements and performing the variations, leads to the following equation for the channel trial functions $F_1$ and $F_2$: 
\be \label{6} 
		F_{1}({\bf q_{1}},-{\bf q_{1}}) \left[ 2\omega({\bf q_{1}},m)\,-\,E \right] 
		= - 4 g_{2} \int {\text d}{\bf p} 
		\frac{F_2({-\bf q_{1}},{\bf q_1}+{\bf p}, {\bf p})}
		{(\sqrt{8 \pi^3})^3\sqrt{2\omega({\bf q_1}, m)}\sqrt{2\omega({\bf p},\mu)}\sqrt{2\omega({\bf q_1}+{\bf p},m)}}
\ee
and	
\be \label{7}
\begin{split}
		 F_{2}({\bf q_1},{\bf q_2},{\bf q_{1}}+{\bf q_{2}}) \; \left[ \omega({\bf q_{1}}, m) + \omega({\bf q_{2}}, m) + \omega({\bf q_{1}}+{\bf q_{2}}, \mu) -\,E\right] \\
		  = - 4 g_{2} \frac{F_1({-\bf q_{2}},{\bf q_{2}})}
  {(\sqrt{(8 \pi^3})^3 \sqrt{2\omega({\bf q_{1}},m)}\sqrt{2\omega({\bf q_{1}}+{\bf q_{2}},\mu)}\sqrt{2\omega(-{\bf q_{2}},m)}}
\end{split}
\ee
\par It is not possible to obtain exact analytic solutions of the coupled, relativistic equations (\ref{6}) and (\ref{7}), so we shall resort to obtaining approximate variational-perturbative solutions. 
\par \ni
In lowest order approximation we take  $~ E ~\simeq  ~\omega({\bf q_1},m) + \omega({\bf q_2},m)~$ 
in equation (\ref{7}), whereupon equation (\ref{7}) simplifies to
\be \label{8}
\begin{split}
	F_{2}({\bf q_1},{\bf q_2},{\bf q_{1}}+{\bf q_{2}}) \left[\omega({\bf q_{1}}+{\bf q_{2}}, \mu)\right] \\
	\,=\,-\, 4 g_{2} \frac{F_1({-\bf q_{2}},{\bf q_{2}})}
	{(\sqrt{(8 \pi^3})^3 \sqrt{2\omega({\bf q_{1}},m)}\sqrt{2\omega({\bf q_{1}}+{\bf q_{2}},\mu)}\sqrt{2\omega(-{\bf q_{2}}, m)}}. 
\end{split}
\ee
Substituting the expression (\ref{8}) into equation (\ref{6}),
 the latter, in the rest frame (i.e. the total momentum ${\bf Q} = 0$), simplifies to the  single relativistic momentum-space equation 
\be \label{9}
	f({\bf q})\left[ 2\omega({\bf q}, m)\,-\,E \right]
	= 4 \, \pi \,  \alpha \,  {m}^2 \int  d^3\,{\bf p}\,
	\frac{f({\bf p})}{\omega({\bf q},m)\,\omega^2({\bf p}-{\bf q},\mu)\,\omega({\bf p},m)}. 
\ee
where $f({\bf q})\,=\,F_1(-{\bf q},{\bf q})\,$, and $\ds \alpha  = \frac{2 g_2^2}{(4 \pi)^4 m^2}$ is a dimensionless coupling constant. The interaction in the integral equation (\ref{9}) is represented by the kernel, which is a relativistic generalization of the potential.  
\par
Our principal interest is to determine the conditions under which the two-Higgs system can form bound states due to a Dark Matter mediating field. For this purpose it is sufficient to study ground states, for which the wave functions are spherically symmetric.
We shall examine the $(\alpha, \mu)$ parameter space, where $ \mu$ is the (unknown) DM particle mass, and $\alpha$ (or, equivalently,  $g_2$), is the (similarly unknown) dimensionless coupling constant. 
\section{Approximate solutions and results}
Unfortunately the relativistic equation (\ref{9}) is not analytically solvable even for spherically symmetric states, so approximate variational solutions will be obtained. 
Variational approximations, as is well known, are only as good as the trial states that are used. 
For our purposes it will be sufficient to use simple ground state trial functions, particularly  in light of the large parameter space to be examined. 
\par
We shall obtain approximate variational solutions of (\ref{9}) for the Higgs-Higgs ground state, using the spherically symmetric trial wave function 
\be \label{10}
\ds f(p) = \frac{\omega(p,\, m)}{(p^2 + b^2)^2} ,
\ee 
where  $p = |\pb|$,  $~ \omega(p, m) = \sqrt{(p^2 + m^2}$ and  $b$ is an adjustable parameter, 
whose value is chosen so that the $E_{{\rm trial}} (b; \alpha, m, \mu)$ is a least upper bound to the unknown 
exact mass of the two-Higgs bound state. 
\par
It should be noted that for non-relativistic Higgs particles, \ie in the limit $\ds \frac{p^2}{m^2} \ll 1$, 
(equivalently, $\omega(p, m) \to m$), equation (\ref{9}) simplifies to 
\be \label{9a}
	f({\bf q})\left[ \frac{\qb^2}{m} - \epsilon \right]
	= 4 \, \pi \,  \alpha \,   \int {\text d}^3\,{\bf p}\,
	\frac{f({\bf p})}{({\bf p}-{\bf q})^2 + \mu^2)},
\ee
where $\epsilon = E - 2 m$. This is recognized to be the momentum space representation of the non-relativistic 
Schr\"odinger equation for the relative motion 
of two Higgs particles interacting via an attractive Yukawa potential $\ds - \, \alpha \frac{e^{- \mu  r}}{r}$.   
As is well known, this non-relativistic equation is not analytically solvable for $\mu >0$ though 
in the limit of massless Dark Matter particles, $\mu = 0$, the solutions are the familiar Hydrogenic ones with
 $\ds \epsilon = - \, \frac{1}{4} \,  m \, \alpha^2$  and $\ds b= \frac{1}{2} m \alpha$ for the ground state. 
\par
The variationally obtained relativistic two-Higgs ground-state rest masses $E$  are listed in Table 1 for various values of the 
dimensionless coupling constant $\alpha$ and for various values of the DM mass $\mu$.  
All results are given in units of the Higgs mass $m$, \ie $E$ is $E/m$, $\mu$ is $\mu / m$, $b$ is $b/m$.

We also list the previously obtained non-relativistic limit results for comparison purposes \cite{14}.
These  relativistic and non relativistic results are also  plotted in { Figure 1} and in { Figure 2}.
\vskip .5cm

%
\vskip .3cm
\ni {\bf{Table 1}}. Variational relativistic two Higgs rest energy $E_{rel}$ for various values of the DM mass $\mu$ (both in units 
of the Higgs mass $m$) and for various values of  the dimensionless coupling constant $\alpha$.  
The quantity $b$ is the optimal value of the variational scale parameter $b$ (in units of the Higgs mass $m$). 
We also give the non-relativistic result $E_{nr}$ for comparison purposes.
\vskip .5cm 
\ni
\begin{tabular}{ |p{2cm}|p{3cm}|p{3cm}|p{3cm}|p{3cm}|  }
 \hline
 \multicolumn{5}{|c|}{$\alpha$ = 0.01} \\
 \hline
$\mu$ & $b$ & $b$ & $E$ & $E$ \\
           & Non-Relativistic & Relativistic & Non-Relativistic & Relativistic \\ 
 \hline
0.0000  &  0.0050000000	&	0.0049989444	&    	1.999975000	&	1.999975000 \\
0.0010  &  0.0048783984        &	0.0048774111         &	1.999983664	&	1.999983666 \\
0.0015  &  0.0047460883	&	0.0047451485	&	1.999987133	&        1.999987136 \\
0.0020  &  0.0045751120        &	0.0045742278	&	1.999990122	&	1.999990123 \\
0.0025  &  0.0043674490 	&        0.0043666272	&	1.999992675	&	1.999992676 \\
0.0030  &  0.0041221774	&	0.0041214241	& 	1.999994832	&	1.999994833 \\
0.0035  &  0.0038348722        &         0.0038341925   	&	1.999996625           &        1.999996626 \\
0.0040  &  0.0034951351 	&         0.0034945326	&	1.999998076	&	1.999998077 \\
0.0045  &  0.0030781877	&	0.0030776632	& 	1.999999201	&	1.999999202 \\					
0.0050  & 0.0025000000         &         0.0024995329        &  	2.000000000	&	unbound \\
 \hline
\end{tabular}
\begin{tabular}{ |p{2cm}|p{3cm}|p{3cm}|p{3cm}|p{3cm}|  }
\hline
\multicolumn{5}{|c|}{$\alpha$ = 0.1} \\
\hline
$\mu$ & $b$ & $b$ & $E$ & $E$ \\
           & Non-Relativistic & Relativistic & Non-Relativistic & Relativistic \\ 
\hline
0.000  & 	0.0500000000	&	0.0490491428	&	 1.997500000	&	1.997527459 \\
0.010  &  	0.0487839844       & 	0.0478935044         &	 1.998366353	&	1.998389559 \\
0.015	 & 	0.0474608825	&	0.0466114592	&	 1.998713376	&	1.998733632 \\
0.020  &  	0.0457511204  	&	0.0449494329	&    	 1.999012120	&	1.999029227 \\ 
0.025	 & 	0.0436744901	&	0.0429265681	&	 1.999267453	&	1.999281386 \\ 
0.030	 & 	0.0412217741	&	0.0405329877	&	 1.999483229          &	1.999494095 \\
0.035	 &  	0.0383487224	&	0.0377237277	&	 1.999662509	&   	1.999670520 \\
0.040	 & 	0.0349513506  	&	0.0343937611	&   	 1.999807643          &  	1.999813089 \\
0.045	 & 	0.0307818766	&	0.0302924647	&	 1.999920155	&	1.999923382 \\
\hline
\end{tabular}

\begin{tabular}{ |p{2cm}|p{3cm}|p{3cm}|p{3cm}|p{3cm}|  }
\hline
\multicolumn{5}{|c|}{$\alpha$ = 0.5} \\
\hline
$\mu$ & $b$ & $b$ & $E$ & $E$ \\
           & Non-Relativistic & Relativistic & Non-Relativistic & Relativistic \\ 
\hline
0.00   &	0.2500000000	&	0.1954804199	&	1.937500000	&	1.947011993 \\
0.01   &	0.2497147851      	&	0.1957478826      	&	1.942353821	&	1.951649179 \\
0.05   &	0.2439199223	&	0.1923172164	&        1.959158829		&	1.967135921 \\
0.1     &	0.2287556021	&	0.1809286299       	&	1.975303007	&	1.981177410 \\
0.15   &	0.2061088708	&	0.1631240265	&	1.987080715	&	1.990833047 \\
0.20   &	0.1747567533	&	0.1375741000	&   	1.995191084	&	1.997083715 \\							
0.21   &	0.1669975210	&	0.1310202512	&      	1.996412850	&	1.997980449 \\
0.23   &  	0.1490363127	&	0.1151845935	&	1.998469730	&	1.999440329 \\
0.24   &  	0.1381835823	&	     &  1.999303047	&	unbound \\
0.245 &	0.1319854886	&	       & 1.999669182	&	unbound \\
\hline
\end{tabular}

\begin{tabular}{ |p{2cm}|p{3cm}|p{3cm}|p{3cm}|p{3cm}|  }
\hline
\multicolumn{5}{|c|}{$\alpha$ = 1.0} \\
\hline
$\mu$ & $b$ & $b$ & $E$ & $E$ \\
           & Non-Relativistic & Relativistic & Non-Relativistic & Relativistic \\ 
\hline
0.0	&	0.5000000000	&	0.3021534764	&	  1.750000000	&			1.828329614 \\
0.01	&	0.4998538419	&	0.3030882883	&          1.759851954	&		 	1.837268511 \\
0.1	&	0.4878398446	&	0.2997899929	&       	  1.836635318	&		 	1.901360074 \\
0.3	&        0.4122177417	&	0.2492925688	&           1.948322860	&		 	1.977746849 \\
0.4	&	0.3495135065	&	0.2026640762	&	   1.980764335	&		 	1.995117073 \\
0.42	&	0.3339950421	&	0.1899549310	&	   1.985651399	&		 	1.997375352 \\
0.45	&	0.3078187666	&	 	 &	   1.992015541	& 	unbound \\
\hline
\end{tabular}

\begin{tabular}{ |p{2cm}|p{3cm}|p{3cm}|p{3cm}|p{3cm}|  }
\hline
\multicolumn{5}{|c|}{$\alpha$ = 1.5} \\
\hline
$\mu$ & $b$ & $b$ & $E$ & $E$ \\
           & Non-Relativistic & Relativistic & Non-Relativistic & Relativistic \\ 
\hline
0.0	&	0.7500000000	&	0.3728822765  	&	1.437500000	&		  1.676269261 \\
0.01	&	0.7499017302	&	0.3742905228  	&	1.452351313	&		  0.372882277 \\	  
0.1	&	0.7413970995	&  	0.3768978968	&	1.573658016	&		  1.785482227 \\
0.3	&	0.6862668064	&  	0.3462987386	&	1.777727066	&		  1.916214889 \\
0.5	&	0.5903671916	&  	0.2821499316	&	1.911494893	&		  1.980371260 \\
0.6	&	0.5242702599	&  	0.2315506632	&	1.956719753	&		  1.996503625 \\
0.7	&	0.4368091778	&  			&	1.988859776	&		  unbound \\
\hline
\end{tabular}

\begin{tabular}{ |p{2cm}|p{3cm}|p{3cm}|p{3cm}|p{3cm}|  }
\hline
\multicolumn{5}{|c|}{$\alpha$ = 2.0} \\
\hline
$\mu$ & $b$ & $b$ & $E$ & $E$ \\
           & Non-Relativistic & Relativistic & Non-Relativistic & Relativistic \\ 
\hline
0.00	&	1.000000000	&	0.4255384887	&	1.000000000	&	1.503833978 \\
0.01	& 	0.9999259798         &	0.4273308572      	&	1.019850988	&	1.520463609 \\		  
0.10	&        0.9933301511         &	0.4341423237      	&	1.185897111	&	1.647178628 \\	  
0.30	&        0.9492176505       	&	0.415127049     	&	1.485350682	&	1.830237683 \\
0.50   &  	0.8734898018       	&	0.3685721652      	&	1.706981212           &	1.932649768 \\      
0.75	&	0.7345173498	&	0.2697740392	&	1.895690531           &	1.994104525 \\      
0.80	&	0.6990270133 	&	0.2369773494      	&	1.923057339	&         1.999960782 \\
0.90	&	0.6156375332	&			& 	1.968062163	&	unbound \\
\hline
\end{tabular}
\newpage

\begin{figure}[!htb]
\centering
\includegraphics[scale=.7]{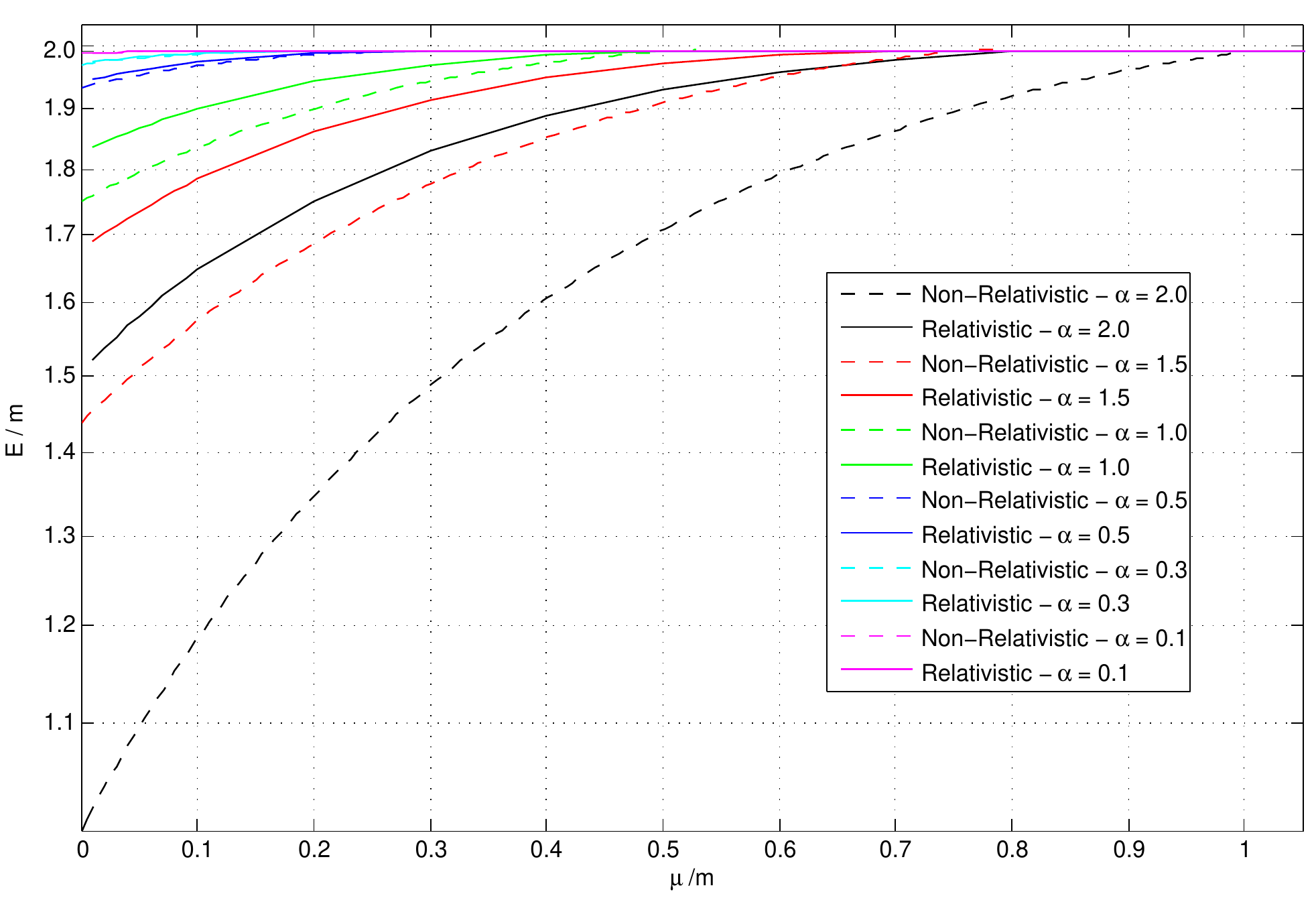}
\caption{The two-Higgs rest mass $E_{min}$ versus DM mass $\mu$, in units of the Higgs mass $m$. Relativistic results are represented by the solid curves, dashed curves represent the corresponding non-relativistic calculations.}
\label{fig:firstgraph}
\end{figure}
\ni
\begin{figure}[!htb]
\centering
\includegraphics[scale=.85]{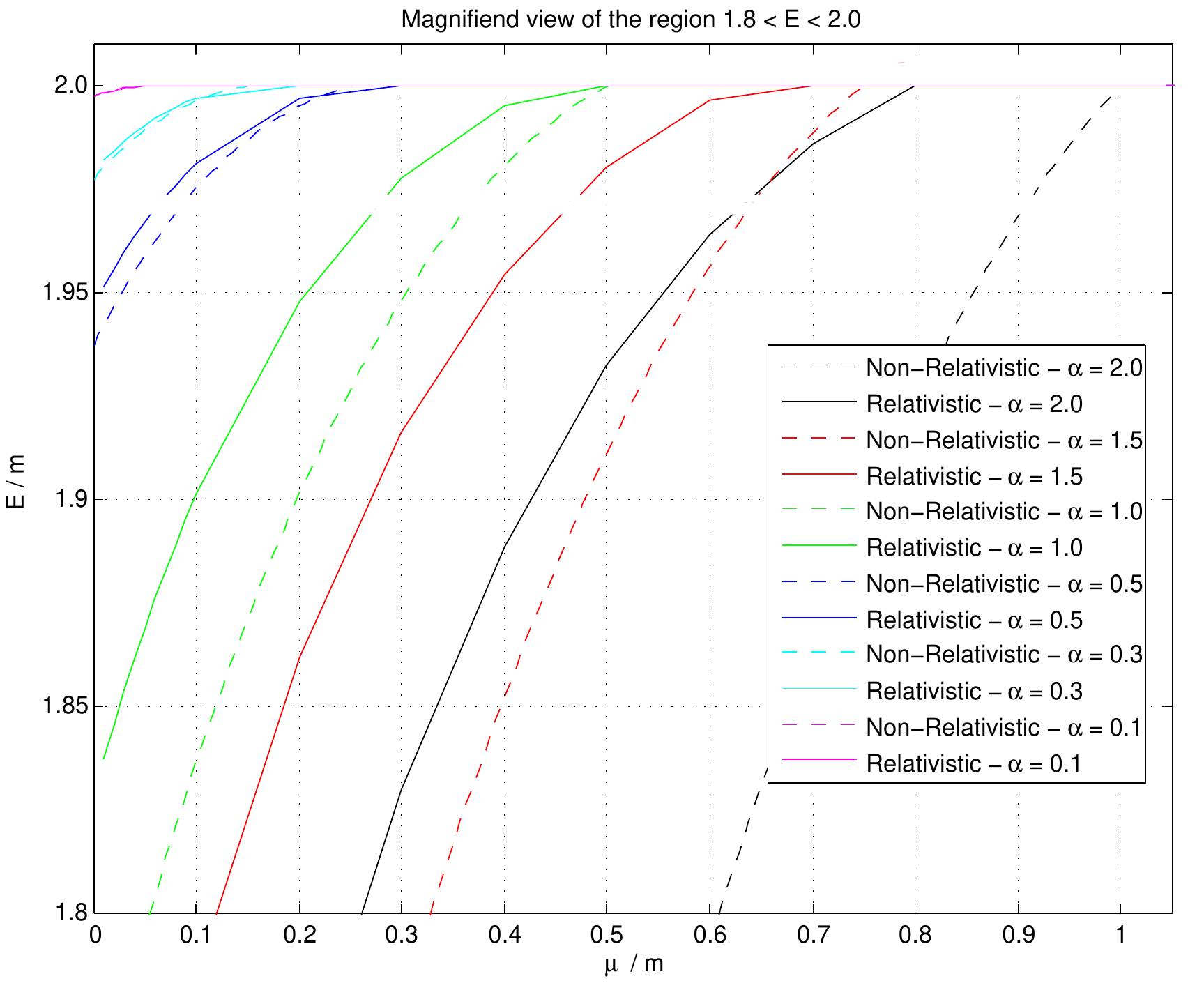}
\caption{Two-Higgs rest mass $E_{min}$ versus DM mass $\mu$, in units of the Higgs mass $m$, 
magnified Figure 1 for the region 1.8 $<$ $E$ $<$ 2.0 }
\label{fig:secondgraph}
\newpage
\end{figure}

\vskip 3mm
\noindent It is evident that the relativistic values of the bound two-Higgs rest mass $E$ are  significantly higher (\ie the binding energy is significantly lower) than the non-relativistic
values, and the difference grows with increasing values of the coupling constant $\alpha$. This underscores  
the importance of using a relativistic description.   
%
\section{Concluding remarks}
Our results show that two-Higgs bound states can be formed due to a massive  spinless real scalar Dark Matter mediating 
field,  over a broad range of the effective dimensionless coupling constant $\alpha=2 g_{2}^{2}/(4\pi)^{4} m^{2}$ and DM mass $\mu$, 
 namely $0 < \alpha \le 2$ and $\mu\, \leq\, m$, i.e. decay of the DM particle into two Higgs particles is not possible. 
\par
Note that the indicated domain of the effective dimentionless coupling constant $\alpha$ does not require that $g_2$  be large 
since $\alpha$ is proportional to $(g_2 / m)^2$.

\par
Of course, these bound states are actually quasi-bound states since the Higgs particle has a very short lifetime 
and so the two-Higgs bound system is also short lived. 
 Such  quasi-bound states are expected to manifest themselves as resonances 
in the scattering cross section in DM on DM collisions. 
This is analogous to bound states of positronium, which are also short-lived quasi-bound states of an 
electron-positron system that manifest themselves as resonances in photon-photon scattering \cite{20}.
\vskip .3cm
In the present work we do not study $H$ on $H$ elastic scattering states, that is solutions of 
equation (\ref {9}) with $E > 2\, m$.  
Also, the effect of the coupling constants other than $g_2$ are not sampled in this work. 
Their effect on the binding energies can be evaluated perturbatively or by expanding the number of Fock components in 
the trial state $|\psi_{trial}\ket$. In any case, their effect would not eliminate the binding, since the interactions in this 
model are overall attractive. 
We shall report on the effect of interaction terms of the Lagrangian (\ref{1.2}) with coupling constants other than $g_2$ 
in subsequent work.

%
%

\end{document}